\documentclass[aps,pra,groupedaddress,floats,reprint]{revtex4-2}

\usepackage{amsfonts,graphicx,soul,natbib,mathrsfs,dsfont,amssymb}

\usepackage{amsmath}
\usepackage{graphicx}
\usepackage{amsthm}
\usepackage{array}
\usepackage{physics}
\usepackage{makecell}
\usepackage[dvipsnames]{xcolor}
\usepackage{colortbl}
\usepackage{mathtools}
\usepackage{stmaryrd}
\usepackage{tcolorbox}
\usepackage{bm}
\usepackage{esvect}
\usepackage[colorlinks=true, allcolors=blue]{hyperref}

\begin{document}

\title{Angular momentum flux of twisted light in paraxial and nonparaxial regimes.}

\author{N.A. Vlasov}%
\email{nikolai.vlasov@metalab.ifmo.ru}
\affiliation{School of Physics and Engineering,
ITMO University, St. Petersburg, Russia 197101}%

\author{N.V. Filina}%
\email{nvfilina@bk.ru}%
\affiliation{School of Physics and Engineering,
ITMO University, St. Petersburg, Russia 197101}%

\author{S.S. Baturin}%
\email{s.s.baturin@gmail.com}%
\affiliation{School of Physics and Engineering,
ITMO University, St. Petersburg, Russia 197101}%
\date{\today}

\date{\today}

\begin{abstract}
We present a theoretical framework for the derivation of the $z$-directed total angular momentum density of a Bessel wave and of the corresponding intercepted angular momentum flux in an ideal perfectly absorbing disk model. Exact expressions for these quantities are obtained in both paraxial and nonparaxial regimes. We then analyze several experimentally relevant setups in the large-argument asymptotic of the Bessel functions. By varying the beam wavelength, polarization, and cone angle, we identify several distinct regimes of the intercepted angular momentum flux. Within the ideal absorber model, this intercepted flux may be identified with the corresponding mechanical torque. The results suggest potential applications for size-sensitive probing and controlled flux manipulation within the idealized macroscopic model.
\end{abstract}


\maketitle

\section{Introduction}

It is well known that electromagnetic waves can carry a spin momentum related to certain circular polarization and orbital angular momentum (OAM) due to its spatial phase distribution. Meanwhile, waves that have an orbital angular momentum and can propagate in a certain direction are of interest. Plane waves can be characterized by a selected propagation direction, but have a zero orbital angular momentum projection along it. Asymmetric spherical waves, on the contrary, have no distinguished radiation direction, but can possess a non-zero projection of orbital angular momentum along any axis. A notable solutions of wave equation such as Bessel, Laguerre-Gaussian and other modes are simultaneous combination of these two properties: a selected propagation direction and a non-zero projection of orbital angular momentum. As a result, wavefront of such waves represents a spiral twisted around the z-axis which indicates direction of radiation. This special state of light is called ``twisted light". 

Early theoretical analyses of electromagnetic fields carrying angular momentum were given by Zel'dovich and co-authors \cite{zeldovich} and, independently, by Vasara and co-authors \cite{vasara}. In these studies the authors discussed electromagnetic field configurations carrying angular momentum. A major step forward was made by J.~F.~Allen and co-authors \cite{allen}, who established the modern paraxial description of laser beams with angular momentum in terms of Laguerre-Gaussian (LG) modes. Within the paraxial approximation, they derived expressions for the angular momentum density and angular momentum flux for arbitrary polarization states, showing that the latter is proportional to $l+\sigma$, where $\sigma=\pm1$ corresponds to right- and left-circular polarization, respectively, and $\sigma=0$ to linear polarization. Subsequently, in Ref.~\cite{nonparaxialAllen}, J.~F.~Allen and co-authors obtained a general nonparaxial expression for the angular momentum per unit length associated with a beam whose electric field is transverse and parametrized by a mode function $E(k)$ and a polarization state. Their result was formulated in terms of integrals over various field components, which must be evaluated for a specific choice of $E(k)$. The authors considered the LG beam as an example and provided an explicit expression for the angular momentum per unit length in the nonparaxial regime.

Beyond Laguerre--Gaussian modes, a variety of other solutions carrying orbital angular momentum have been investigated. In Ref.~\cite{sztul}, Airy-beam solutions of the paraxial wave equation were considered, and the corresponding Poynting vector and angular momentum density were constructed numerically. For Bessel beams, the total angular momentum density was derived in Ref.~\cite{volke} using a vector spherical harmonics approach. Superpositions of Bessel and Bessel--Gauss beams were studied in Refs.~\cite{litvin,dudley}, where expressions for the total angular momentum density were obtained for linearly polarized fields, and experimental measurements of the angular momentum density were reported. In Ref.~\cite{Jaimes} authors constructed Hankel–Laguerre waves solutions to paraxial wave equations and, moreover, proved that all structured wavefields described by the paraxial wave equation diverge for large radial distances. In Ref.~\cite{Rodriguez} the solution for scalar non-paraxial wave equation was reported in terms of spheroidal beams. The complete description of accelerating traveling waves with parabolic rotational symmetry has
been presented in Ref.~\cite{Espindola}.

Experimental demonstrations have confirmed that electromagnetic waves can transfer not only linear momentum but also angular momentum to matter. At the microscopic scale, absorbing microparticles placed in focused laser beams were observed to undergo rotational motion determined by the handedness of the twisted light \cite{arita}. At the macroscopic scale, a suspended quarter-wave plate in vacuum was shown to rotate under the action of circularly polarized light \cite{beth}. These experiments provided direct evidence of angular momentum transfer from light to matter. Owing to these fundamental demonstrations and the associated control capabilities, twisted light has remained an active area of research, with applications ranging from optical manipulation and trapping of micro-objects to broader photonic and technological contexts \cite{soifer}.

Despite extensive theoretical and experimental efforts, a closed-form expression for the intercepted $z$-directed angular momentum flux of a Bessel beam on a finite absorbing disk, valid for arbitrary polarization states and arbitrary paraxiality, is still lacking. In this work, we address this problem within an ideal perfectly absorbing thin-disk model. More precisely, we derive exact expressions for the total angular momentum density of a twisted electromagnetic Bessel wave and for the corresponding intercepted angular momentum flux, which in the ideal absorber model is identified with the mechanical torque exerted on the disk. We emphasize that the present work does not attempt a microscopic description of light-matter transfer. In general, such a description requires an explicit material-response model and may involve induced currents, polarization, reflection, scattering, and partial absorption; see, e.g., Refs.~\cite{Babiker, Maslov}. Within the idealized macroscopic model considered here, we analyze several distinct regimes that arise when varying the beam wavelength, polarization, and cone angle. We further discuss how these regimes may be accessed in principle and how the corresponding intercepted angular momentum flux can be controlled. All calculations are performed in SI units, and whenever plotted quantities depend on the omitted field amplitude they are shown in a normalized form.

The present work builds on earlier results for angular momentum in paraxial and nonparaxial structured beams, including field-based treatments of Bessel and related modes~\cite{allen,nonparaxialAllen,volke,litvin,dudley}, and extends them to the finite-size interception problem considered here. Specifically, we derive explicit helicity-resolved expressions for electromagnetic Bessel waves in the representation adopted here and use them to obtain a closed-form expression for the intercepted $z$-directed total angular momentum flux on a finite disk for arbitrary helicity and arbitrary paraxiality within the ideal absorber model. We further develop a systematic asymptotic comparison of the paraxial and strongly nonparaxial regimes. Based on the derived formulas, we propose several potential applications in both the paraxial and nonparaxial regimes. We show that varying the degree of paraxiality allows one to control the beam wavelength and the characteristic size of objects for which the proposed effects can be observed.

\section{4-potential of twisted light}

This section describes the derivation of the 4-potential
$A^{\mu} = \left(A^0 (\mathbf{r}, t), \mathbf{A}(\mathbf{r}, t)\right)^T$
of Bessel waves corresponding to twisted photons. We begin our analysis in the Coulomb gauge,
$A^0 (\mathbf{r}, t)=0$ and $\nabla \cdot \mathbf{A}(\mathbf{r}, t)=0$.
Next, we restrict ourselves to monochromatic waves of the form
$\mathbf{A}(\mathbf{r}, t) = \mathbf{A}(\mathbf{r}) e^{-i \omega t}$.
Under these assumptions, the problem reduces to determining the spatial amplitude
$\mathbf{A}(\mathbf{r})$ satisfying the Helmholtz equation
\begin{equation}
\label{eq:Helmholz}
     \left[\nabla^2 + k^2 \right] \mathbf{A} (\mathbf{r}) = 0,
\end{equation}
where $k=\omega/c$ is the wavenumber.

In Ref.~\cite{serbo}, the authors propose a clear and elegant approach for constructing solutions to Eq.~(\ref{eq:Helmholz}) describing Bessel waves of twisted photons. Their idea is to represent the vector potential as a coherent superposition of vector plane waves,
\begin{equation}
\label{eq:superposition}
     \mathbf{A}_{\Lambda} (\mathbf{r}) =
     \int a_{k_r \ell}(\mathbf{\varkappa}) \,
     \mathbf{A}_{\mathbf{k} \Lambda} (\mathbf{r}) \,
     \dfrac{d^2 \bm{\varkappa}}{(2\pi)^2},
\end{equation}
where the total wavenumber is defined as $k = \sqrt{k_z^2 + k_r^2}$.
The coefficient
$a_{k_{r} \ell}(\mathbf{\varkappa}) = i^{-\ell} \exp(i \ell \varphi)\,
\dfrac{2 \pi}{\varkappa} \delta(\varkappa - k_r)$
is the Fourier amplitude, and
$\mathbf{A}_{\mathbf{k} \Lambda} (\mathbf{r}) =
\mathbf{e}_{\mathbf{k} \Lambda} \exp(i \mathbf{k}\mathbf{r})$
is the vector potential of a plane wave with helicity $\Lambda = \pm 1$.
Here, $\mathbf{e}_{\mathbf{k} \Lambda}$ is the photon polarization vector, which is an eigenvector of the helicity operator
$\hat{\Lambda} = \hat{\mathbf{s}} \cdot \mathbf{k} / k$,
\begin{equation}
\label{eq:helicity}
     \hat{\Lambda} \mathbf{e}_{\mathbf{k} \Lambda} = \Lambda \mathbf{e}_{\mathbf{k} \Lambda}.
\end{equation}
It should be noted that a given value of the photon helicity $\Lambda$ corresponds to a specific circular polarization state of the beam.

After a series of algebraic transformations, the authors of Ref.~\cite{serbo} obtained the following expression for the vector potential of a Bessel beam of twisted light:
\begin{equation}
\label{eq:vecpoten}
    \mathbf{A}_{\Lambda} (\mathbf{r}) =
    \sum_{\sigma=0, \pm 1}
    i^{-\sigma} d^1_{\sigma \Lambda}(\theta)
    J_{l-\sigma} (k_r r)
    e^{i (l-\sigma) \varphi} e^{i k_z z}
    \bm{\chi}_{\sigma},
\end{equation}
where $\theta = \arcsin(k_r/k)$ is the opening angle of the beam and
$d^1_{\sigma \Lambda}(\theta)$ are the Wigner small-$d$ matrices,
$d^1_{\Lambda, \Lambda} = \cos^2(\theta/2)$,
$d^1_{-\Lambda, \Lambda} = \sin^2(\theta/2)$, and
$d^1_{0, \Lambda} = \Lambda \sin{\theta} / \sqrt{2}$. Above we assumed the unit amplitude factor 
$A_0 = 1 \; T \cdot m$. 
The vectors $\bm{\chi}_{\sigma}$ form the spiral basis and satisfy the orthonormality relations
\[
\bm{\chi_{\sigma}^{*}} \bm{\chi_{\sigma'}} = \delta_{\sigma \sigma'},
\]
with explicit forms
\[
\bm{\chi}_{\pm} = \mp \dfrac{1}{\sqrt{2}}
\begin{pmatrix}
  1 \\
  \pm i \\
  0
\end{pmatrix},
\qquad
\bm{\chi}_{0} =
\begin{pmatrix}
  0 \\
  0 \\
  1
\end{pmatrix}.
\]

For our purposes, it is more convenient to work in the polar basis.
The polar basis vectors are related to the spiral basis as follows (see Appendix~\ref{Change basis}):
\begin{equation}
\begin{cases}
    \mathbf{e}_r =
    \dfrac{1}{\sqrt{2}}
    \left(\bm{\chi}_- e^{i \varphi} - \bm{\chi}_+ e^{-i \varphi}\right), \\
    \mathbf{e}_{\varphi} =
    \dfrac{i}{\sqrt{2}}
    \left(\bm{\chi}_- e^{i \varphi} + \bm{\chi}_+ e^{-i \varphi}\right), \\
    \mathbf{e}_z = \mathbf{e}_z.
\end{cases}
\end{equation}

After the transformations described in Appendix~\ref{Change basis}, we obtain the coordinate-dependent components of the vector potential in the polar basis:
\begin{multline}
    A_{r} =
    \bra{\mathbf{e}_{r}}\ket{\mathbf{A}_{\Lambda} (\mathbf{r})} =
    \dfrac{i}{2\sqrt{2}} e^{i l \varphi} e^{i k_z z} \times \\
    \times \left[(1-\Lambda \cos{\theta}) J_{l+1}
    + (1+\Lambda \cos{\theta}) J_{l-1} \right],
\end{multline}
\begin{multline}
    A_{\varphi} =
    \bra{\mathbf{e}_{\varphi}}\ket{\mathbf{A}_{\Lambda} (\mathbf{r})} =
    \dfrac{1}{2 \sqrt{2}} e^{i l \varphi} e^{i k_z z} \times \\
    \times \left[(1-\Lambda \cos{\theta}) J_{l+1}
    - (1+\Lambda \cos{\theta}) J_{l-1} \right],
\end{multline}
\begin{equation}
    A_{z} =
    \bra{\mathbf{e}_{z}}\ket{\mathbf{A}_{\Lambda} (\mathbf{r})} =
    \dfrac{1}{\sqrt{2}} e^{i l \varphi} e^{i k_z z}
    \Lambda \sin{\theta} J_l .
\end{equation}

Here and in what follows, for compactness we introduce the notation
$J_l \equiv J_l(k_r r)$.

\section{Electromagnetic field and energy distribution of twisted light}

Since we have factored out the time dependence $e^{-i \omega t}$ in the vector potential, the electric and magnetic fields are assumed to have the same harmonic time dependence:
$\mathbf{B}(\mathbf{r}, t) = \mathbf{B}(\mathbf{r}) e^{-i \omega t}$ and
$\mathbf{E}(\mathbf{r}, t) = \mathbf{E}(\mathbf{r}) e^{-i \omega t}$.
Therefore, it is sufficient to evaluate only the coordinate-dependent parts of the fields.

We begin with the magnetic field, which is obtained from the vector potential as
\begin{equation}
\label{eq:B evaluation}
     \mathbf{B} (\mathbf{r}, t) = \nabla \times \mathbf{A} (\mathbf{r}, t)
\end{equation}
After the derivation presented in Appendix~\ref{magnetic field evaluation}, we obtain the following expression for the magnetic field of twisted light:
\begin{multline}
\label{eq:magnetic field-2}
     \mathbf{B} (\mathbf{r}, t) = \dfrac{k \Lambda e^{i l \varphi} e^{i k_z z} e^{-i \omega t}}{2 \sqrt{2}} \times \\
     \times \begin{pmatrix}
     i \left[ (1 - \Lambda \cos{\theta}) J_{l+1} + (1 + \Lambda\cos{\theta}) J_{l-1}  \right] \\
     \left[ (1 - \Lambda \cos{\theta}) J_{l+1} - (1 + \Lambda\cos{\theta}) J_{l-1}  \right] \\
     2 \Lambda \sin{\theta} J_l
\end{pmatrix}
\end{multline}

The electric field in the Coulomb gauge is evaluated as
\begin{equation}
\label{eq:electic eq 1}
     \mathbf{E} (\mathbf{r}, t) = - \dfrac{\partial \mathbf{A} (\mathbf{r}, t)}{\partial t} = i ck \mathbf{A} (\mathbf{r}) e^{-i \omega t}
\end{equation}
Therefore, we obtain the electric field of twisted light in the form
\begin{multline}
\label{eq:electric field}
     \mathbf{E} (\mathbf{r}, t) = \dfrac{ick e^{i l \varphi} e^{i k_z z} e^{-i \omega t}}{2 \sqrt{2}} \times \\
     \times \begin{pmatrix}
     i \left[ (1 - \Lambda \cos{\theta}) J_{l+1} + (1 + \Lambda\cos{\theta}) J_{l-1}  \right] \\
     \left[ (1 - \Lambda \cos{\theta}) J_{l+1} - (1 + \Lambda\cos{\theta}) J_{l-1}  \right] \\
     2 \Lambda \sin{\theta} J_l
\end{pmatrix}
\end{multline}

The Poynting vector is conventionally defined as
\begin{equation}
\label{eq:poyting vector}
    \mathbf{S} = \dfrac{1}{4 \mu_0}
    \left[\mathbf{E} \times \mathbf{B}^* + \mathbf{E}^* \times \mathbf{B}\right].
\end{equation}

\begin{figure}
\centering
\includegraphics[width = \columnwidth]{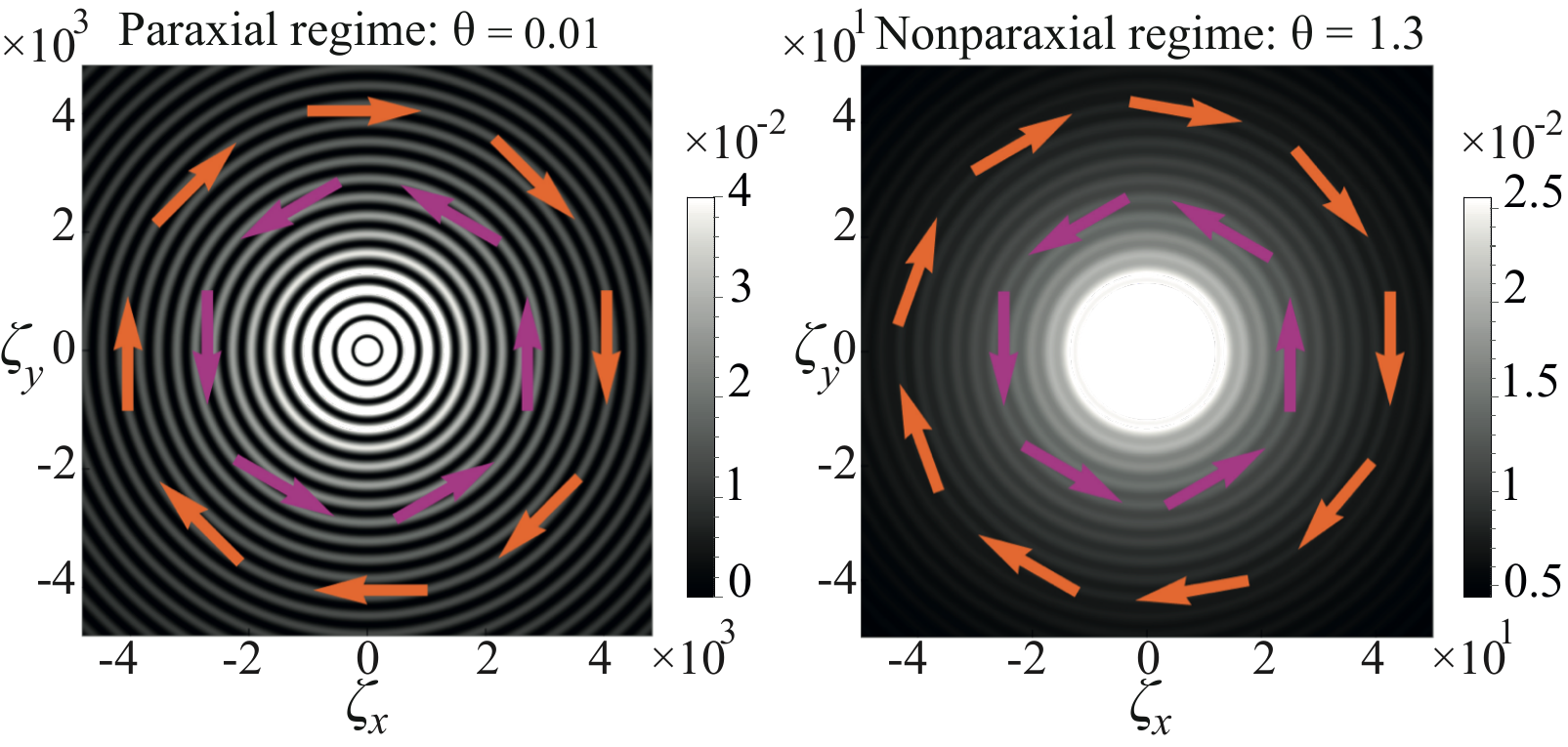} 
\caption{Normalized energy density $\tilde{u}$ and transverse Poynting vector $\tilde{\mathbf{S}}_{\perp}$ for twisted radiation in the paraxial regime (left) and the nonparaxial regime (right). The beam parameters $l=1$ and $\Lambda=1$ are used. Orange arrows correspond to clockwise circulation of the Poynting vector, while purple arrows correspond to counterclockwise circulation.}
\label{E}
\end{figure}

The azimuthal component $S_{\varphi}$ is given by
\begin{equation}
\label{eq:sphi}
S_{\varphi} =
\dfrac{1}{4 \mu_0}
\left(
E_z B_r ^{*} - E_r B_z ^{*}
+ E_z ^{*} B_r - E_r ^{*} B_z
\right).
\end{equation}

Substituting the expressions for the electric and magnetic fields into Eq.~(\ref{eq:sphi}), we obtain (see Appendix~\ref{Poyting evaluation} for details)
\begin{equation}
    S_{\varphi} =
    \dfrac{c k^2}{2 \mu_0}
    \left[
    \dfrac{l}{k r} \left( 1 + \Lambda \cos{\theta} \right) J_l ^2
    -\dfrac{\Lambda}{2} \sin{2 \theta} J_l J_{l+1}
    \right].
\end{equation}

Since it is straightforward to verify that $S_r = 0$, the dimensionless transverse Poynting vector can be written as
\begin{equation}
    \tilde{\mathbf{S}}_{\perp} =
    \dfrac{2 \mu_0}{c k^2}
    \begin{pmatrix}
        0 \\
        S_{\varphi}
    \end{pmatrix}.
\end{equation}

As expressions for the fields and the relevant components of the Poynting vector are now available, we can visualize the spatial distribution of the energy density and the Poynting vector in a plane perpendicular to the beam propagation axis.

We define a dimensionless energy density using the normalized electric and magnetic fields,
$\tilde{\mathbf{E}} = \mathbf{E}/(k A_0)$ and
$\tilde{\mathbf{B}} = \mathbf{B}/(k A_0)$, as
\begin{equation}
    \tilde{u} =
    \dfrac{|\tilde{\mathbf{E}}|^2/c^2 + |\tilde{\mathbf{B}}|^2}{2}.
\end{equation}

We plot the dimensionless energy density and the dimensionless Poynting vector in a plane perpendicular to the propagation axis using dimensionless coordinates $(\zeta, \varphi)$, where $\zeta = kr$, for a beam with $l=1$ and right-handed circular polarization $\Lambda=1$ (see Figure~\ref{E}). Both the paraxial regime, $\theta = 0.01~\mathrm{rad}$, and the nonparaxial regime, $\theta = 1.3~\mathrm{rad}$, are considered. In the paraxial regime, oscillations of the dimensionless energy density appear at significantly larger values of $\zeta$ than in the nonparaxial regime, namely at $\zeta \gtrsim 10^3$ versus $\zeta \gtrsim 10^1$, respectively. This behavior follows from the asymptotic condition for Bessel functions, $k_r r \gg |l^2 - 1/4|$, which can be rewritten as $\zeta \gg |l^2 - 1/4|/\sin{\theta}$. In the nonparaxial limit $\theta \to \pi/2$, this condition is satisfied at much smaller values of $\zeta$, whereas in the paraxial limit $\theta \to 0$ it requires much larger values of $\zeta$. In addition, the amplitude of energy-density oscillations decreases more rapidly with increasing $\zeta$ in the nonparaxial regime. In both regimes, the Poynting vector describes a circular energy flow.

\section{Angular-momentum flux-density pseudotensor of twisted light}

Next, we are interested in determining the total angular momentum flux carried by the twisted beam. But, before calculating the total angular momentum flux, we first evaluate the angular-momentum flux-density pseudotensor~\cite{Novotny}. The angular-momentum flux-density pseudotensor is defined as
\begin{equation}
    \bm{\stackrel{\leftrightarrow}{\rho}} = \mathbf{\stackrel{\leftrightarrow}{T}} \times \mathbf{r},
\end{equation}
where Maxwell's stress tensor in vacuum is
\begin{equation}
    \mathbf{\stackrel{\leftrightarrow}{T}} = \varepsilon_0 \mathbf{E} \mathbf{E} + \frac{1}{\mu_0} \mathbf{B} \mathbf{B} - \frac{1}{2} \left(\varepsilon_0 E^2 + \frac{1}{\mu_0} B^2\right) \mathbf{\stackrel{\leftrightarrow}{I}}.
\end{equation}
Thus, the $zz$ component of the time-averaged angular-momentum flux-density pseudotensor, which corresponds to the total time-averaged angular momentum flux along the beam propagation axis, is given by
\begin{equation}
    \langle\rho_{zz}\rangle = -r \langle T_{z\varphi} \rangle,
\end{equation}
with time-averaged Maxwell's stress tensor component
\begin{equation}
    \langle T_{z\varphi} \rangle = \frac{1}{2} \Re\left[\varepsilon_0 E_z E^*_\varphi + \frac{1}{\mu_0} B_z B^*_\varphi\right].
\end{equation}
Using the obtained formulas for electromagnetic field of twisted light, we now obtain the following expression for the time-averaged angular-momentum flux-density:
\begin{equation}
    \langle\rho_{zz}\rangle =
    \dfrac{k A_0^2}{2 \mu_0}
    \left[
    l \left( \cos{\theta} + \Lambda \right) J_l ^2
    -\Lambda kr \sin{\theta} J_l J_{l+1}
    \right].
\end{equation}

We next consider the asymptotic behavior of the angular momentum flux-density in the limit
$k_r r \gg |l^2 - 1/4|$, which yields

\begin{multline}
    \langle\rho_{zz}\rangle(k_r r \gg |l^2 - 1/4|) \to
    \dfrac{k A_0^2}{2 \mu_0 \pi}
    \Bigg\{
    \Lambda \cos{(2k_r r - \pi l)} \\
    + \dfrac{l}{k r \sin{\theta}} \left[\cos{\theta} - \dfrac{\Lambda}{2} \right] + \dfrac{\sin{(2k_r r - \pi l)}}{k r \sin{\theta}} \\
    \times \left[ l \cos{\theta} - \Lambda\left(l^2 + \dfrac{1}{4}\right)\right] \Bigg\}.
\end{multline}

Thus, in the limit $k_r r \gg |l^2 - 1/4|$, the angular momentum flux-density becomes bounded from above and below. The upper bound is

\begin{equation}
\label{up_rho}
    \overline{\langle\rho_{zz}\rangle}
    =
    \dfrac{k A_0^2}{2 \mu_0  \pi},
\end{equation}
while the lower bound is
\begin{equation}
\label{down_rho}
    \underline{\langle\rho_{zz}\rangle}
    =
    -\dfrac{k A_0^2}{2 \mu_0  \pi}.
\end{equation}
As can be seen, the amplitude of the angular momentum flux-density oscillations depends on the wavenumber $k$.

\begin{figure}
\centering
\includegraphics[width = \columnwidth]{ 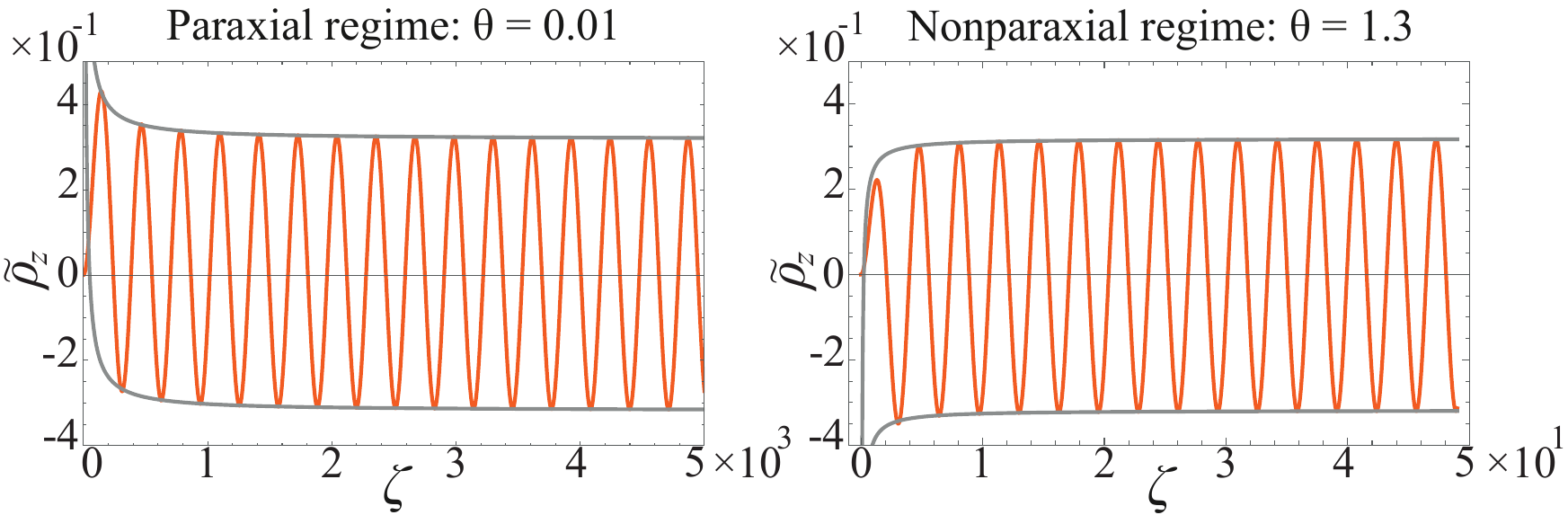} 
\caption{Dependence of the dimensionless angular momentum density $\tilde{\rho}_z$ on the dimensionless parameter $\zeta = kr$ at fixed wavenumber $k$. The left panel corresponds to the paraxial regime with $\theta = 0.01$, while the right panel corresponds to the nonparaxial regime with $\theta = 1.3$. The gray lines indicate the upper and lower bounds of the angular momentum density in the asymptotic regime $\zeta \sin{\theta} \gg |l^2 - 1/4|$.}
\label{rho(x)}
\end{figure}

Finally, we plot the dimensionless angular momentum flux-density
$\langle\tilde{\rho}_{zz}\rangle = 2 \langle\rho_{zz}\rangle \mu_0 / (k A^2_0)$
as a function of the dimensionless parameter $\zeta = kr$ at fixed $k$ (see Figure~\ref{rho(x)}):

\begin{multline}
    \langle\tilde{\rho}_{zz}\rangle (\zeta) =
    l \left( \cos{\theta} + \Lambda \right)
    J_l^2 (\zeta \sin{\theta}) \\
    - \Lambda
    \zeta \sin{\theta}
    J_l (\zeta \sin{\theta})
    J_{l+1} (\zeta \sin{\theta}).
\end{multline}

Both the paraxial regime, $\theta = 0.01~\mathrm{rad}$, and the nonparaxial regime, $\theta = 1.3~\mathrm{rad}$, are shown. In the nonparaxial regime, oscillations of the angular momentum flux-density appear at smaller values of $\zeta$ than in the paraxial regime, which is explained by the same asymptotic condition for Bessel functions as in the case of the energy density. In the paraxial regime, oscillations start at $\zeta \gtrsim 10^3$, whereas in the nonparaxial regime they appear already at $\zeta \gtrsim 10^1$. Moreover, the oscillation amplitude is larger in the paraxial regime. In both cases, the angular momentum flux-density approaches a purely oscillatory behavior described by $\cos{(2 \zeta \sin{\theta} - \pi l)}$ with fixed amplitude in the large-$\zeta$ limit (see Eqs.~\eqref{up_rho} and \eqref{down_rho}), even though the energy density oscillates with a decreasing amplitude. This purely oscillatory asymptotic behavior is related to infinite beam  energy and circular polarization, as it is proportional to $\Lambda$. In contrast, in Ref.~\cite{dudley}, where linear polarization was considered, attenuation of the angular momentum density was observed. Thus, describing twisted beams in the circular-polarization basis reveals a qualitatively different behavior of the angular momentum flux-density.

\section{Total angular momentum flux}

We calculate the total angular momentum flux of the twisted light as an integral of the flux-density of the angular momentum over the cross section with radius $a$ (which may correspond to radius of the irradiated disk what we discuss further in applications):

\begin{equation}
    M_z = -  \int \langle\bm{\stackrel{\leftrightarrow}{\rho}}\rangle \cdot \; \mathbf{n}(\mathbf{r}) ds =
    \int\limits_0^a \int\limits_0^{2\pi} r \langle\rho_{zz}\rangle \mathrm{d} r \mathrm{d} \varphi,
\end{equation}
where $\mathbf{n}(\mathbf{r}) = -\mathbf{e}_z$ is the unit vector perpendicular to the disk, and $ds$ is
an infinitesimal surface element.

After rendering the integral dimensionless, we obtain
\begin{multline}
    M_z = \dfrac{\pi}{\mu_0 k \sin^2{\theta}}
    \left[
    l \cos{\theta}\int\limits_0^{k_r a} J^2_l(y) \, y \mathrm{d} y
    \right. \\
    \left.
    + \Lambda 
    \int\limits_0^{k_r a} J_l (y) \dot{J}_l (y) \, y^2 \mathrm{d} y
    \right].
\end{multline}

Introducing the notation $x = k a$, we obtain the final expression for the $z$ projection of the total angular momentum flux (see Appendix~\ref{total momentum} for a detailed derivation):
\begin{multline}
\label{TAM}
M_z =
\dfrac{a \pi}{2 \mu_0} x
\left[
l \cos{\theta} J^2_l(x \sin{\theta})
\right. \\
\left.
- (l \cos{\theta}-\Lambda) J_{l+1} (x \sin{\theta}) J_{l-1} (x \sin{\theta})
\right].
\end{multline}

We now consider the limit $x \sin{\theta} \gg |l^2 - 1/4|$, which yields
\begin{multline}
M_z (x \sin\theta \gg |l^2 - 1/4|) \to  \\
\dfrac{a}{\mu_0 \sin{\theta}}
\left[
l \cos{\theta} - \Lambda
\sin^2{\left(x \sin{\theta} - \dfrac{\pi l}{2} - \dfrac{\pi}{4}\right)}
\right].
\end{multline}

The upper bound for the total angular momentum flux is
\begin{equation}
\overline{M_z} =
\left\{
\begin{aligned}
    &\dfrac{a l \cos{\theta}}{\mu_0 \sin{\theta}},
    && \text{if } \Lambda = 1, \\
    &\dfrac{a}{\mu_0 \sin{\theta}} (l \cos{\theta} + 1),
    && \text{if } \Lambda = -1 .
\end{aligned}
\right.
\end{equation}

The lower bound is
\begin{equation}
\underline{M_z} =
\left\{
\begin{aligned}
    &\dfrac{a}{\mu_0 \sin{\theta}} (l \cos{\theta} - 1),
    && \text{if } \Lambda = 1, \\
    &\dfrac{a l \cos{\theta}}{\mu_0 \sin{\theta}},
    && \text{if } \Lambda = -1 .
\end{aligned}
\right.
\end{equation}

Our results indicate that the total angular momentum flux is proportional to the vortex charge $l$ and exhibits oscillatory behavior as a function of the dimensionless parameter $x$. The oscillation pattern depends on the helicity $\Lambda$ and on the degree of paraxiality, while the common prefactor scales as $\sin^{-1}{\theta}$.

Since the total angular momentum flux depends explicitly on the paraxiality degree, we now examine two limiting regimes in more detail: the paraxial and nonparaxial regimes.

\subsection{Paraxial regime}

In the paraxial regime, we assume $\sin{\theta} \approx \theta$ and $\cos{\theta} \approx 1$.
The expression for the total angular momentum flux then becomes
\begin{multline}
\label{M_paraxial}
M_z (\cos{\theta} \approx 1) =
\dfrac{a \pi}{2 \mu_0} x
\left[
l J^2_l(x \theta)
\right. \\
\left.
- (l-\Lambda) J_{l+1} (x \theta) J_{l-1} (x \theta)
\right].
\end{multline}

In the limit $x \theta \gg |l^2 - 1/4|$, we obtain
\begin{multline}
M_z (\cos{\theta} \approx 1,\; x \theta \gg |l^2 - 1/4|) \to \\
\dfrac{a}{\mu_0 \theta}
\left[
l - \Lambda
\sin^2{\left(x \theta - \dfrac{\pi l}{2} - \dfrac{\pi}{4}\right)}
\right].
\end{multline}

The upper bound for the total angular momentum flux is
\begin{equation}
\overline{M_z} =
\left\{
\begin{aligned}
    &\dfrac{a l}{\mu_0 \theta},
    && \text{if } \Lambda = 1, \\
    &\dfrac{a}{\mu_0 \theta} (l + 1),
    && \text{if } \Lambda = -1 .
\end{aligned}
\right.
\end{equation}

The lower bound is
\begin{equation}
\underline{M_z} =
\left\{
\begin{aligned}
    &\dfrac{a}{\mu_0 \theta} (l-1),
    && \text{if } \Lambda = 1, \\
    &\dfrac{a l}{\mu_0 \theta},
    && \text{if } \Lambda = -1 .
\end{aligned}
\right.
\end{equation}

\subsection{Nonparaxial regime}

In the nonparaxial regime, we assume $\sin{\theta} \approx 1$ and
$\cos{\theta} \approx \pi/2 - \theta  = \delta \ll 1$.
Under these approximations, the expression for the $z$ projection of the total angular momentum flux takes the form
\begin{multline}
\label{nonparaxial tam}
M_z (\cos{\theta} \approx \pi/2 - \theta) = \\
\dfrac{a \pi}{2 \mu_0} x
\left[
l \delta J^2_l(x)
- (l \delta - \Lambda) J_{l+1} (x) J_{l-1} (x)
\right].
\end{multline}

In the asymptotic regime $x \gg |l^2 - 1/4|$, we obtain
\begin{multline}
M_z (\cos{\theta} \approx \pi/2 - \theta,\; x \gg |l^2 - 1/4|) \to \\
\dfrac{a}{\mu_0}
\left[
l \delta - \Lambda
\sin^2{\left(x - \dfrac{\pi l}{2} - \dfrac{\pi}{4}\right)}
\right].
\end{multline}

The upper bound for the total angular momentum flux is
\begin{equation}
\overline{M_z} =
\left\{
\begin{aligned}
    &\dfrac{a l \delta}{\mu_0},
    && \text{if } \Lambda = 1, \\
    &\dfrac{a}{\mu_0} (l \delta + 1),
    && \text{if } \Lambda = -1 .
\end{aligned}
\right.
\end{equation}

The lower bound is
\begin{equation}
\underline{M_z} =
\left\{
\begin{aligned}
    &\dfrac{a}{\mu_0} (l \delta - 1),
    && \text{if } \Lambda = 1, \\
    &\dfrac{a l \delta}{\mu_0},
    && \text{if } \Lambda = -1 .
\end{aligned}
\right.
\end{equation}

Our results indicate that in the strongly nonparaxial regime $\delta \ll |l|^{-1}$ the total angular momentum flux ceases to be proportional vortex charge $l$, but the amplitude is totally determined by helicity $\Lambda$ and dimensionless parameter $x$. In this regime the vortex charge $l$ determines only oscillation period of total angular momentum flux.

\section{Applications}

Having obtained exact expressions for the total angular momentum flux in both the paraxial and nonparaxial regimes, we can now discuss several potential applications and notable features that emerge in these two limits. 

We first note the physical limitations of our model. The regimes discussed below rely on the large-argument asymptotics of the Bessel functions, $k_\perp a \gg 1$. In the strongly nonparaxial case, $k_\perp \sim k$, this implies that the disk radius must satisfy $\lambda \ll a$. Furthermore, the ideal Bessel beam is an infinite-energy solution and should be viewed as the limiting case of a finite-aperture quasi-Bessel or Bessel--Gauss beam. The present results are expected to remain applicable provided the beam envelope varies weakly across the object, i.e.\ $a \ll w$, where $w$ denotes the transverse envelope scale. Throughout this section, we therefore adopt an ideal perfectly absorbing disk model, in which the intercepted $z$-directed angular momentum flux may be identified with the mechanical torque exerted on the object. This approximation does not constitute a microscopic theory of absorption: in general, angular momentum transfer to matter depends on the detailed material response and may involve reflection, scattering, and partial absorption. We are fully aware of these complications, and for explicit light-matter treatments and possible extensions in this direction we refer the reader to Refs.~\cite{Babiker,Maslov}.

To relate the amplitude factor $A_0$ to a measurable beam characteristic, one may replace the ideal Bessel beam by a finite-energy Bessel--Gauss regularization,
\begin{align}
\mathbf{A}^{(\mathrm{BG})}_{\Lambda}(\mathbf{r})
=
A_0\,e^{-r^2/w^2}\,\mathbf{A}_{\Lambda}(\mathbf{r}),
\end{align}
where $w$ is the transverse envelope scale. The total incident power is then defined by the axial Poynting flux,
\begin{align}
\label{eq:Power}
    P = 2\pi \int_{0}^{\infty}  S^{(\mathrm{BG})}_{z}(r) rdr
    \equiv A_0^2\,\mathcal{P}_{l,\Lambda}(k_r,\theta,w),
\end{align}
with
\begin{align}
    A_0=\sqrt{\frac{P}{\mathcal{P}_{l,\Lambda}(k_r,\theta,w)}} .
\end{align}
Thus, all normalized intercepted-flux curves shown below can be converted to physical units by multiplying them by the factor $A_0^2=P/\mathcal{P}_{l,\Lambda}$. In the broad-envelope regime $a\ll w$, this regularization does not change the leading local behavior over the disk.

At the level of scaling, since $\mathbf{E}\sim c k \mathbf{A}$ and $\mathbf{B}\sim k \mathbf{A}$, the axial Poynting flux of the regularized beam behaves as
\begin{equation}
\label{eq:S_z}
S^{(\mathrm{BG})}_z(r)
\sim
\frac{c k^2}{\mu_0}\,A_0^2\,e^{-2r^2/w^2}\,
f_{l,\Lambda}(k_r r,\theta),
\end{equation}
where $f_{l,\Lambda}$ is a dimensionless mode-shape factor of order unity in the bright part of the beam. Consequently,
\begin{align}
P
\sim
\frac{\pi c k^2 w^2}{2\mu_0}\,A_0^2\,\eta_{l,\Lambda}(\theta),
\end{align}
with $\eta_{l,\Lambda}(\theta)$ a dimensionless factor set by the detailed mode profile.

Hence,
\begin{align}
\label{eq:A0}
A_0 \sim \sqrt{\frac{2\mu_0\,P}{\pi c k^2 w^2\,\eta_{l,\Lambda}(\theta)}}.
\end{align}
Therefore, all normalized curves shown below can be converted to physical units by multiplying them by the factor $A_0^2$ in the units of $(T\cdot m)^2$. In Appendix \ref{app:sec:E} we present exact expression for vector potential amplitude \eqref{eq:A0} for Bessel-Gauss regularization. For example, for the incident paraxial beam with an opening angle $\theta = 0.01$ rad, wavelength $\lambda = 532$ nm, polarization $\Lambda = 1$, $l=1$, power $1$ kW, and the transverse envelope scale $w = 5$ cm the vector potential amplitude takes the value of
$A_0 \approx 3.4 \cdot 10^{-10}$ T $\cdot$ m.

\subsection{Manipulation of the intercepted angular momentum flux by tuning the wavelength}

One experimentally relevant setting corresponds to objects of fixed size irradiated by twisted light. In our notation, this implies that the parameter $a$ is fixed. In this case, the expressions derived for the $z$ projection of the total angular momentum flux allow one to vary parameters associated solely with the properties of the incident beam. In particular, the wavenumber $k$, which is related to the wavelength by $k = 2\pi/\lambda$, can be tuned. Varying the wavelength of twisted light therefore provides a direct means of manipulating the intercepted angular momentum flux. Below, we highlight several notable regimes illustrating this control.

Examining the expression for the total angular momentum flux in both paraxial and nonparaxial regimes, Eq.~(\ref{TAM}), we identify a particularly simple and illustrative case for 
$l=\Lambda=\pm1$. 

In the paraxial regime the expression for the total angular momentum flux then reads (see Eq.~(\ref{M_paraxial}))
\begin{equation}
M_z (\cos{\theta} \approx 1, l=\Lambda=\pm1)
= \pm \dfrac{a \pi}{2 \mu_0} x J^2_1 (x \theta).
\end{equation}

In the asymptotic regime $x \theta \gg 3/4$, this expression reduces to
\begin{equation}
M_z (x \theta \gg 3/4) \to
\pm \dfrac{a}{\mu_0 \theta}
\sin^2{\left(x \theta - \dfrac{\pi}{4}\right)}.
\end{equation}

In strongly nonparaxial regime, $\delta \ll |l|^{-1}$, the expression for total angular momentum flux (see Eq.~(\ref{nonparaxial tam})) in this case reads
\begin{multline}
M_z(\cos{\theta} \approx 0, l=\Lambda=\pm1) = \pm
\dfrac{a \pi}{2 \mu_0} x J_2 (x) J_0 (x).
\end{multline}

In the asymptotic regime $x \gg 15/4$, this expression reduces to
\begin{align}
M_z (x \theta \gg 3/4) \to
\mp
\dfrac{a}{\mu_0} \cos^2{\left(x-\frac{\pi}{4}\right)}.
\end{align}

This regime is notable in that the angular momentum flux is entirely determined by oscillations of the squared Bessel function. As a result, by tuning the wavelength $\lambda$, one can reach values at which the total angular momentum flux vanishes. In this way, the intercepted angular momentum flux can be effectively switched on and off by varying the wavelength. Moreover, changing the circular polarization reverses the sign of the total angular momentum flux, providing, within the ideal absorber model, direct control over the sign of the intercepted angular momentum flux and hence over the sign of the corresponding mechanical torque. It is worth to note that this effect can be achieved for strongly nonparaxial beams with any value of $l$ as in this regime total angular momentum ceases to be proptional $l$, while for nonparaxial beams this works only if $l=1$. The only dependence on $l$ in nonparaxial regime is in the period of oscillations and in the visibility limits of this effect since it is originates from asymptotic regime $\lambda \lesssim2 \pi a/|l^2 - 1/4|$.

\begin{figure}[t]
\centering
\includegraphics[width = \columnwidth]{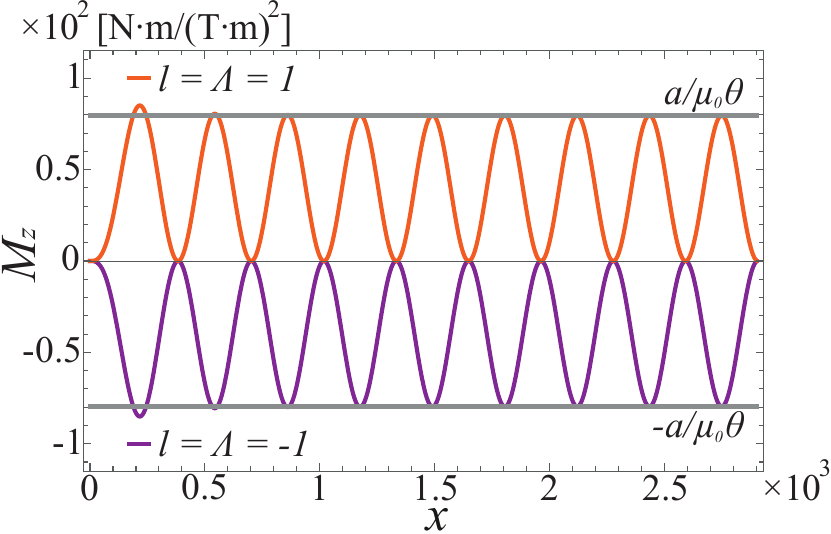}
\caption{$z$ projection of the total angular momentum flux in the paraxial regime, $\theta = 0.01$, for two beams with $l=\Lambda=1$ (orange line) and $l=\Lambda=-1$ (purple line) , normalized by the squared amplitude of the vector potential $A_0$. In both cases the object size is fixed. The gray lines indicate the asymptotic bounds of the total angular momentum flux, $\pm a/(\mu_0\theta)$.}
\label{onoff}
\end{figure}

Figure~\ref{onoff} shows the dependence of the total angular momentum flux in the paraxial ($\theta = 0.01$) regime on the dimensionless parameter $x$ for beams with $l=\Lambda=1$ and $l=\Lambda=-1$, assuming a fixed object size $a=1~\mu\mathrm{m}$. The behavior is symmetric for the two polarizations, and oscillations appear only for $x \gtrsim 10^3$. This implies that, for an object of micrometer size, the wavelength must satisfy $\lambda \lesssim 2\pi a \times 10^{-3}$ (which corresponds to soft X-ray/EUV range) in order for these effects to be observable.

\begin{figure}[t]
\centering
\includegraphics[width = \columnwidth]{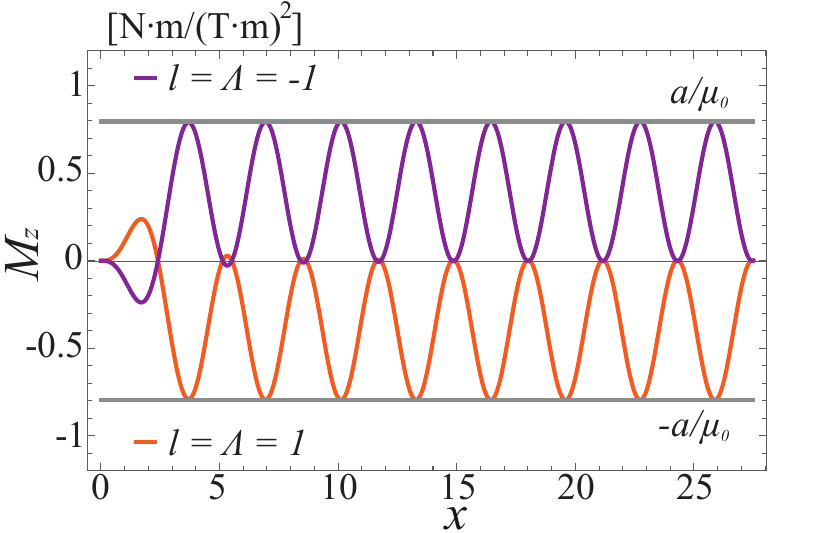}
\caption{$z$ projection of the total angular momentum flux in the nonparaxial regime for two beams with $l=\Lambda=1$ (orange line) and $l=\Lambda=-1$ (purple line) , normalized by the squared amplitude of the vector potential $A_0$. In both cases the object size is fixed. The gray lines indicate the asymptotic bounds of the total angular momentum flux, $\pm a/\mu_0$.}
\label{onoff_nonparaxial}
\end{figure}

Figure~\ref{onoff_nonparaxial} shows the dependence of the total angular momentum flux in the strongly nonparaxial ($\delta \ll |l|^{-1}$) regime on the dimensionless parameter $x$ for beams with $l=\Lambda=1$ and $l=\Lambda=-1$, assuming a fixed object size $a=1~\mu\mathrm{m}$. The behavior is symmetric for the two polarizations, and oscillations appear only for $x \gtrsim 10$. This implies that, for an object of micrometer size, the wavelength must satisfy $\lambda \lesssim 2\pi a \times 10^{-1}$ (which corresponds to optical/UV range) in order for these effects to be observable. Increasing value of $l$ blue-shifts the visibility limit of this effect in nonparaxial regime. Moreover, we note that in nonparaxial regime the sign of total angular momentum flux reverses for fixed polarization.

\subsection{Distinguishing the size of objects}

If the radiation wavelength $\lambda$ is fixed, the expression for the total angular momentum flux in the paraxial approximation becomes
\begin{align}
&M_z (\cos{\theta} \approx 1) = \\ \nonumber
&\dfrac{\pi}{2 \mu_0 k} x^2
\left[
l J^2_l(x \theta)
- (l-\Lambda) J_{l+1} (x \theta) J_{l-1} (x \theta)\right].
\end{align}

In the asymptotic limit $x \theta \gg |l^2 - 1/4|$, this reduces to
\begin{align}
&M_z (\cos{\theta} \approx 1,\; x \theta \gg |l^2 - 1/4|) \to \\ \nonumber
&\dfrac{x}{\mu_0 \theta k}
\left[
l - \Lambda
\sin^2{\left(x \theta - \dfrac{\pi l}{2} - \dfrac{\pi}{4}\right)}
\right].
\end{align}

Considering the case $l=1$ and $\Lambda=\pm1$, we obtain
\begin{multline}
M_z (\cos{\theta} \approx 1,\; x \theta \gg |l^2 - 1/4|,\; l=1,\; \Lambda=\pm1) \to \\
\dfrac{x}{\mu_0 \theta k}
\left[
1 \mp \cos^2{\left(x \theta - \dfrac{\pi}{4}\right)}
\right].
\end{multline}

Considering strongly nonparaxial regime, $\delta \ll |l|^{-1}$, for fixed radiation wavelength $\lambda$ the epxression for the total angular momentum flux reads

\begin{equation}
M_z(\cos{\theta} \approx 0) = \pm
\dfrac{x^2 \pi}{2 \mu_0 k} J_{l+1} (x) J_{l-1} (x).
\end{equation}

In the asymptotic regime $x \gg |l^2 - 1/4|$, this expression reduces to
\begin{align}
M_z (\cos{\theta} \approx 0, x \gg |l^2 - 1/4|) \to
\mp
\dfrac{x}{\mu_0 k} \sin^2{\left(x-\frac{\pi l}{2} - \frac{\pi}{4}\right)}.
\end{align}

Considering the case $l=1$ and $\Lambda=\pm1$, we obtain
\begin{align}
M_z (\cos{\theta} \approx 0,\; x \gg |l^2 - 1/4|,\; l=1,\; \Lambda=\pm1) \to \\ \nonumber
\mp \dfrac{x}{\mu_0 k} \cos^2{\left(x-\frac{\pi}{4}\right)}.
\end{align}

In paraxial regime, at points $x \theta = 3\pi/4 + \pi n$, with $n \in \mathbb{Z}$, the values of $M_z$ for $\Lambda=\pm1$ coincide. In contrast, at points $x \theta = \pi/4 + \pi n$ the difference between two polarizations is maximal: for $\Lambda=-1$ the total angular momentum flux reaches its maximum value, while for $\Lambda=1$ it vanishes. In strongly nonparaxial regime, at points $x = 3\pi/4 + \pi n$, with $n \in \mathbb{Z}$, the values of $M_z$ for $\Lambda=\pm1$ mutually equal zero, while at points $x \theta = \pi/4 + \pi n$ the difference between  two polarizations is maximal. Thus, at a fixed wavelength, changing the polarization enables discrimination between objects of different sizes in both paraxial and nonparaxial regimes.

\begin{figure}[t]
\centering
\includegraphics[width = \columnwidth]{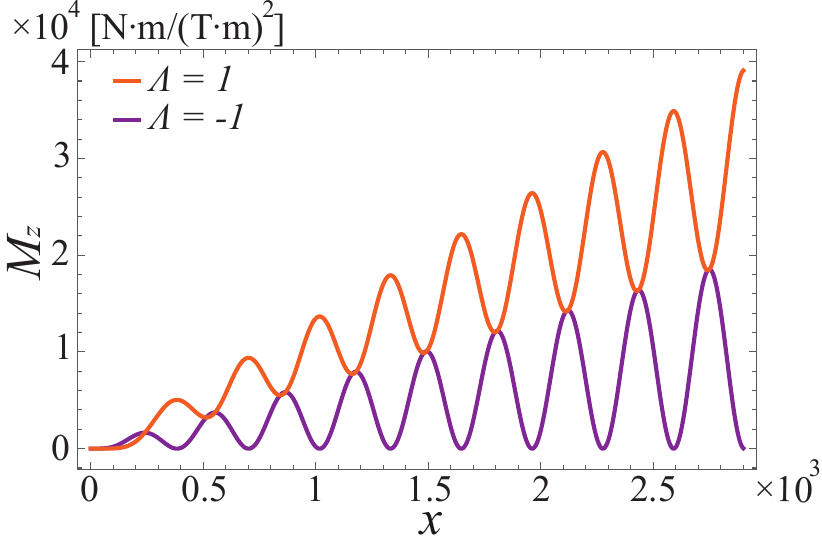}
\caption{$z$ projection of the total angular momentum flux in the paraxial regime for right-circularly polarized ($\Lambda=1$, orange line) and left-circularly polarized ($\Lambda=-1$, purple line) beams with $l=1$, normalized by the squared amplitude of the vector potential $A_0$. The radiation wavelength is fixed.}
\label{sizes}
\end{figure}

Figure~\ref{sizes} illustrates this behavior for right- and left-circularly polarized beams with $l=1$ in paraxial regime ($\theta=0.01$) at a fixed wavelength $\lambda=532~\mathrm{nm}$. We can note that for these parameters the observable object's sizes are $a \gtrsim 40$ micron. For distinguishing micron-scale objects ($a \sim 1$ micron) one needs soft X-ray/EUV range lasers.

\begin{figure}[t]
\centering
\includegraphics[width = \columnwidth]{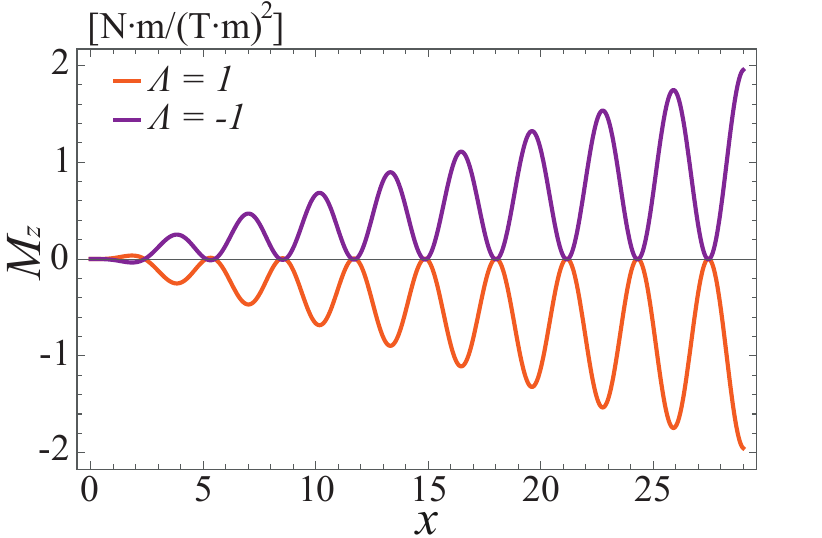}
\caption{$z$ projection of the total angular momentum flux in the nonparaxial regime for right-circularly polarized ($\Lambda=1$, orange line) and left-circularly polarized ($\Lambda=-1$, purple line) beams with $l=1$, normalized by the squared amplitude of the vector potential $A_0$. The radiation wavelength is fixed.}
\label{sizes_nonparaxial}
\end{figure}

Figure~\ref{sizes_nonparaxial} illustrates the object's discrimination for right- and left-circularly polarized beams with $l=1$ in strongly nonparaxial regime ($\delta \ll |l|^{-1}$) at a fixed wavelength $\lambda=532~\mathrm{nm}$. We can note that for these parameters the observable object's sizes are $a \gtrsim 0.85$ micron. Moreover, it is worth to note that the object's discrimination regime is observable for any value of $l$, since $l$ determines only low boundary of observable object's sizes at fixed wavelength from asymptotic condition $a \gtrsim \lambda|l^2 - 1/4|/2\pi$.

\section{Conclusion}

In this study, we developed a unified field-based framework for the total angular momentum density of twisted electromagnetic Bessel waves and for the corresponding intercepted total angular momentum flux on a finite disk within the ideal perfectly absorbing model.
Starting with an exact representation of electromagnetic fields in the Coulomb gauge, we derived closed-form expressions for total angular momentum density and the intercepted angular momentum flux. These results are valid for arbitrary degrees of paraxiality and general polarization states. From a fundamental standpoint, we demonstrate that the total angular momentum flux exhibits oscillations as a function of the dimensionless parameter $x=ka$ when intercepted by an object of finite size. Moreover, in the paraxial regime, the amplitude of the total angular momentum flux is proportional to both the vortex charge $l$ and the helicity $\Lambda$. In contrast, in the strongly nonparaxial regime, it becomes proportional only to $\Lambda$, while the dependence on $l$ manifests solely in the oscillation period.

Based on these exact expressions, we identified several experimentally relevant regimes of intercepted angular momentum flux. First, for a fixed object size, the intercepted angular momentum flux can be efficiently controlled by varying the beam wavelength, which directly determines the dimensionless parameter $x = ka$. For a fixed object size, increasing the degree of nonparaxiality allows one to blue-shift the corresponding wavelengths at which this effect can be observed. Second, we demonstrated that the intercepted angular momentum flux exhibits strong sensitivity to the helicity of the incident light at specific values of $x$ and cone angle $\theta$. Switching the circular polarization of the beam can maximize or suppress the intercepted flux, which may provide a mechanism for distinguishing objects of different sizes under identical illumination conditions within the idealized model. For a fixed wavelength, increasing the degree of nonparaxiality enables the distinguishing of smaller objects compared to the paraxial regime.

The results presented here clarify how the intercepted angular momentum flux of twisted light behaves beyond the paraxial approximation and establish a direct link between nonparaxial field geometry, polarization, and finite-size effects. From a practical standpoint, the demonstrated sensitivity of the intercepted angular momentum flux to beam parameters and object size suggests potential applications in size-selective probing and in controlled manipulation of the intercepted angular momentum flux within the idealized absorber model. In the perspective, our analysis should extend naturally from a single Bessel OAM component to composed/engineered illuminations with a prescribed spiral spectrum, i.e., a superposition (or mixture) of OAM channels which may be useful in distinguishing objects with different shapes~\cite{Torres, Molina, Molina1, Torner}.

\begin{acknowledgments}

The authors are greatful to G.K. Syzikh, E. Kim and Yu. Shulakova for many fruitfull discussions. The work of N.A.V. was supported by the by Russian Science
Foundation, Grant No. 25-12-00213.

\end{acknowledgments}

\appendix
\onecolumngrid

\section{Transition from spiral basis to polar basis}
\label{Change basis}

The polar basis is related to the Cartesian basis as follows:
\begin{equation}
    \begin{cases}
    \mathbf{e}_{r} = \mathbf{e}_x \cos{\varphi} + \mathbf{e}_y \sin{\varphi}, \\
    \mathbf{e}_{\varphi} = - \mathbf{e}_x \sin{\varphi} + \mathbf{e}_y \cos{\varphi}, \\
    \mathbf{e}_z = \mathbf{e}_z .
    \end{cases}
\end{equation}

In turn, the Cartesian basis is related to the spiral basis in the following way:
\begin{equation}
    \begin{cases}
        \mathbf{e}_x = -\dfrac{1}{\sqrt{2}} (\bm{\chi}_+ - \bm{\chi}_{-}), \\
        \mathbf{e}_y = -\dfrac{1}{\sqrt{2} i} (\bm{\chi}_+ + \bm{\chi}_{-}), \\
        \mathbf{e}_z = \mathbf{e}_z .
    \end{cases}
\end{equation}

Combining these relations, the polar basis vectors can be expressed in terms of the spiral basis as follows:
\begin{multline*}
\mathbf{e}_r =
- \dfrac{1}{\sqrt{2}} \cos{\varphi} (\bm{\chi}_+ - \bm{\chi}_{-})
- \dfrac{1}{\sqrt{2} i} \sin{\varphi} (\bm{\chi}_+ + \bm{\chi}_-) \\
= - \dfrac{1}{\sqrt{2}}
\left[
\dfrac{e^{i \varphi} + e^{-i \varphi}}{2} (\bm{\chi}_+ - \bm{\chi}_-)
- \dfrac{e^{i \varphi} - e^{-i \varphi}}{2} (\bm{\chi}_+ + \bm{\chi}_-)
\right]
= \dfrac{1}{\sqrt{2}} (\bm{\chi}_- e^{i \varphi} - \bm{\chi}_+ e^{-i \varphi}),
\end{multline*}

\begin{multline*}
\mathbf{e}_{\varphi} =
\dfrac{1}{\sqrt{2}} \sin{\varphi} (\bm{\chi}_+ - \bm{\chi}_{-})
- \dfrac{1}{\sqrt{2} i} \cos{\varphi} (\bm{\chi}_+ + \bm{\chi}_-) \\
=
\dfrac{1}{\sqrt{2} i}
\left[
\dfrac{e^{i \varphi} - e^{-i \varphi}}{2} (\bm{\chi}_+ - \bm{\chi}_-)
- \dfrac{e^{i \varphi} + e^{-i \varphi}}{2} (\bm{\chi}_+ + \bm{\chi}_-)
\right]
= \dfrac{i}{\sqrt{2}} (\bm{\chi}_- e^{i \varphi} + \bm{\chi}_+ e^{-i \varphi}),
\end{multline*}

\begin{flushleft}
$\quad \mathbf{e}_z = \mathbf{e}_z$.
\end{flushleft}

The general expression for the vector potential of a Bessel beam of twisted light in the spiral basis is given by Eq.~\eqref{eq:vecpoten}:
\begin{equation}
\label{eq:vecpotapp}
    \mathbf{A}_{\Lambda} (\mathbf{r}) =
    \sum_{\sigma=0, \pm 1}
    i^{-\sigma}
    d^1_{\sigma \Lambda}(\theta)
    J_{l-\sigma} (k_r r)
    e^{i (l-\sigma) \varphi}
    e^{i k_z z}
    \bm{\chi}_{\sigma}.
\end{equation}

The Wigner small $d$-matrices $d^1_{\sigma \Lambda} (\theta)$ are defined according to Ref.~\cite{serbo} as
\begin{equation}
    d^1_{\Lambda \Lambda}(\theta) = \cos^2{(\theta/2)}, \quad
    d^1_{-\Lambda \Lambda}(\theta) = \sin^2{(\theta/2)}, \quad
    d^1_{0 \Lambda}(\theta) = \Lambda \sin{\theta} / \sqrt{2}.
\end{equation}
Accordingly, the functions $d^1_{1 \Lambda} (\theta)$ and $d^1_{-1 \Lambda} (\theta)$ can be written in the form
\begin{equation}
    d^1_{1 \Lambda} (\theta) = \frac{1}{2} (1 + \Lambda \cos{\theta}), \quad
    d^1_{-1 \Lambda} (\theta) = \frac{1}{2} (1 - \Lambda \cos{\theta}).
\end{equation}

To facilitate subsequent calculations, we express each component of the vector potential in the spiral basis, Eq.~\eqref{eq:vecpotapp}, in terms of the Bessel function $J_l \equiv J_l(k_r r)$ and its derivative with respect to the argument,
$\dot{J}_l \equiv \left. \frac{\mathrm{d} J(x)}{\mathrm{d} x} \right|_{x = k_r r}$:
\begin{align}
    &A_{z} =
    \bra{\mathbf{\chi}_{0}}\ket{\mathbf{A}_{\Lambda} (\mathbf{r})}
    = d^1_{0 \Lambda} J_l e^{i l \varphi} e^{i k_z z}, \\
    &A_{+} =
    \bra{\mathbf{\chi}_{+}}\ket{\mathbf{A}_{\Lambda} (\mathbf{r})}
    = - i d^1_{1 \Lambda} J_{l-1} e^{i(l-1) \varphi} e^{i k_z z}
    = -i \dfrac{d^1_{1 \Lambda}}{d^1_{0 \Lambda}} e^{-i \varphi}
    \frac{J_{l-1}}{J_l} A_z \nonumber\\
    &\hspace{1.8cm}
    = -i \dfrac{d^1_{1 \Lambda}}{d^1_{0 \Lambda}} e^{-i \varphi}
    \left(\dfrac{l}{k_r r} + \dfrac{\dot{J}_l}{J_l}\right) A_z, \\
    &A_{-} =
    \bra{\mathbf{\chi}_{-}}\ket{\mathbf{A}_{\Lambda} (\mathbf{r})}
    = i d^1_{-1 \Lambda} J_{l+1} e^{i(l+1) \varphi} e^{i k_z z}
    = i \dfrac{d^1_{-1 \Lambda}}{d^1_{0 \Lambda}} e^{i \varphi}
    \dfrac{J_{l+1}}{J_l} A_z \nonumber\\
    &\hspace{1.8cm}
    = i \dfrac{d^1_{-1 \Lambda}}{d^1_{0 \Lambda}} e^{i \varphi}
    \left(\dfrac{l}{k_r r} - \dfrac{\dot{J}_l}{J_l}\right) A_z .
\end{align}

Here we have used the recurrence relations for Bessel functions with shifted indices,
\begin{align}
    J_{l-1} (x) = \dfrac{l}{x} J_l(x) + \dot{J}_l(x), \\
    J_{l+1}(x) = \dfrac{l}{x} J_l(x) - \dot{J}_l(x).
\end{align}

Finally, the components of the vector potential in the polar basis are given by
\begin{align}
    &A_{z} =
    \bra{\mathbf{e}_{z}}\ket{\mathbf{A}_{\Lambda} (\mathbf{r})}
    = d^1_{0 \Lambda} J_l e^{i l \varphi} e^{i k_z z}, \\
    &A_{r} =
    \bra{\mathbf{e}_{r}}\ket{\mathbf{A}_{\Lambda} (\mathbf{r})}
    = \dfrac{i}{\sqrt{2}} \dfrac{A_z}{d^1_{0 \Lambda}}
    \left[
    d^1_{1 \Lambda} \left(\dfrac{l}{k_r r} + \dfrac{\dot{J}_l}{J_l}\right)
    + d^1_{-1 \Lambda} \left(\dfrac{l}{k_r r} - \dfrac{\dot{J}_l}{J_l}\right)
    \right] \nonumber\\
    &\hspace{1.8cm}
    = \dfrac{i}{\sqrt{2}} \dfrac{A_z}{d^1_{0 \Lambda}}
    \left(
    \dfrac{l}{k_r r} + \Lambda \cos{\theta}
    \dfrac{\dot{J}_l}{J_l}
    \right), \\
    &A_{\varphi} =
    \bra{\mathbf{e}_{\varphi}}\ket{\mathbf{A}_{\Lambda} (\mathbf{r})}
    = -\dfrac{1}{\sqrt{2}} \dfrac{A_z}{d^1_{0 \Lambda}}
    \left[
    d^1_{1 \Lambda} \left(\dfrac{l}{k_r r} + \dfrac{\dot{J}_l}{J_l}\right)
    - d^1_{-1 \Lambda} \left(\dfrac{l}{k_r r} - \dfrac{\dot{J}_l}{J_l}\right)
    \right] \nonumber\\
    &\hspace{1.8cm}
    = -\dfrac{1}{\sqrt{2}} \dfrac{A_z}{d^1_{0 \Lambda}}
    \left(
    \Lambda \cos{\theta} \dfrac{l}{k_r r}
    + \dfrac{\dot{J}_l}{J_l}
    \right).
\end{align}

\section{Evaluation of magnetic field}
\label{magnetic field evaluation}

The magnetic field is related to the vector potential through the relation
\begin{equation}
\label{eq:Bapp}
     \mathbf{B} (\mathbf{r}) = \nabla \times \mathbf{A} (\mathbf{r}). 
\end{equation}

In polar coordinates, the curl operator takes the form
\begin{align}
    \nabla \times \mathbf{A} =
    \mathbf{e}_{r}
    \left(
    \dfrac{1}{r} \dfrac{\partial A_z}{\partial \varphi}
    - \dfrac{\partial A_\varphi}{\partial z}
    \right)
    + \mathbf{e}_{\varphi}
    \left(
    \dfrac{\partial A_r}{\partial z}
    - \dfrac{\partial A_z}{\partial r}
    \right)
    + \mathbf{e}_{z}
    \dfrac{1}{r}
    \left[
    \dfrac{\partial (r A_\varphi)}{\partial r}
    - \dfrac{\partial A_r}{\partial \varphi}
    \right].
\end{align}

After substituting the vector potential expressed in polar coordinates into Eq.~\eqref{eq:Bapp}, we obtain the following expressions for the components of the magnetic field:
\begin{align}
    B_{r} &=
    \dfrac{il}{r}A_z - ik_z A_{\varphi}
    = i A_z
    \left[
    \dfrac{l}{r}
    + \dfrac{k_z}{\sqrt{2} d^1_{0 \Lambda}}
    \left(
    \Lambda \cos{\theta}
    \dfrac{l}{k_r r}
    + \dfrac{\dot{J}_l}{J_l}
    \right)
    \right] \nonumber \\
    &=
    i A_z
    \left[
    \dfrac{l}{r}
    + \dfrac{k \cos{\theta}}{\Lambda \sin{\theta}}
    \left(
    \Lambda \cos{\theta}
    \dfrac{l}{k \sin{\theta} \, r}
    + \dfrac{\dot{J}_l}{J_l}
    \right)
    \right]
    = \nonumber \\
    &=
    \dfrac{i k A_z}{J_l \sin{\theta}}
    \left(
    \dfrac{l}{k_r r} J_l
    + \Lambda \cos{\theta} \dot{J}_l
    \right)
    =
    \dfrac{i k A_z}{2 J_l \sin{\theta}}
    \left[
    (1 - \Lambda \cos{\theta}) J_{l+1}
    + (1 + \Lambda\cos{\theta}) J_{l-1}
    \right],
\end{align}
\begin{align}
    B_{\varphi} &=
    i k_z A_r
    - k_r d^1_{0 \Lambda} \dot{J}_l (k_r r)
    e^{i l \varphi} e^{i k_z z}
    = \nonumber \\
    &=
    - A_z
    \left[
    \dfrac{k \cos{\theta}}{\Lambda \sin{\theta}}
    \left(
    \dfrac{l}{k_r r}
    + \Lambda \cos{\theta}
    \dfrac{\dot{J}_l}{J_l}
    \right)
    + k \sin{\theta}
    \dfrac{\dot{J}_l}{J_l}
    \right]
    = \nonumber \\
    &=
    - \dfrac{A_z k}{J_l \sin{\theta}}
    \left(
    \Lambda \cos{\theta}
    \dfrac{l}{k_r r} J_l
    + \dot{J}_l
    \right)
    =
    \dfrac{A_z k}{2 J_l \sin{\theta}}
    \left[
    (1 - \Lambda \cos{\theta}) J_{l+1}
    - (1 + \Lambda\cos{\theta}) J_{l-1}
    \right], \\
    \nonumber \\
    B_{z} &=
    \dfrac{A_{\varphi}}{r}
    + \dfrac{\partial A_{\varphi}}{\partial r}
    - il \dfrac{A_r}{r}
    = \nonumber \\
    &=
    \dfrac{A_{\varphi}}{r}
    + k_r \dfrac{\dot{J}_l}{J_l} A_{\varphi}
    + \dfrac{A_z}{\sqrt{2} d^1_{0 \Lambda}}
    \left(
    \Lambda \cos{\theta}
    \dfrac{l}{k_r r^2}
    - \dfrac{\ddot{J}_l J_l - \dot{J}^2_l}{J^2_l} k_r
    \right)
    - \dfrac{il}{r} A_r
    = \nonumber \\
    &=
    - \dfrac{A_z}{\sqrt{2} d^1_{0 \Lambda}}
    \Bigg[
    \dfrac{1}{r} \dfrac{\dot{J}_l}{J_l}
    + k_r \left(\dfrac{\dot{J}_l}{J_l}\right)^2
    + \dfrac{\ddot{J}_l J_l - \dot{J}^2_l}{J^2_l} k_r
    - \dfrac{l^2}{k_r r^2}
    \Bigg]
    = \nonumber \\
    &=
    - \dfrac{A_z k_r}{\sqrt{2} d^1_{0 \Lambda}}
    \left[
    \dfrac{1}{k_r r} \dfrac{\dot{J}_l}{J_l}
    + \dfrac{\ddot{J}_l}{J_l}
    - \dfrac{l^2}{(k_r r)^2}
    \right]
    = \nonumber \\
    &=
    - \dfrac{A_z k_r}{\sqrt{2} d^1_{0 \Lambda}}
    \left[
    \dfrac{l^2}{(k_r r)^2}
    - 1
    - \dfrac{l^2}{(k_r r)^2}
    \right]
    =
    \dfrac{A_z k_r}{\sqrt{2} d^1_{0 \Lambda}}
    = A_z k \Lambda .
\end{align}

\section{Evaluation of 
$\varphi$-component of Poynting vector}
\label{Poyting evaluation}

The electric field in the Coulomb gauge can be obtained from the vector potential using Eq.~\eqref{eq:electic eq 1},
\begin{equation}
     \mathbf{E} (\mathbf{r}, t) =
     - \dfrac{\partial \mathbf{A} (\mathbf{r}, t)}{\partial t}
     = i c k \mathbf{A} (\mathbf{r}) e^{-i \omega t}.
\end{equation}

The $\varphi$ component of the Poynting vector is given by
\begin{equation}
    S_{\varphi} =
    \dfrac{1}{2 \mu_0}
    \Re \left(
    E_z B_r ^{*} - E_r B_z ^{*}
    \right).
\end{equation}

Below we provide a step-by-step evaluation of $S_{\varphi}$:
\begin{align}
    E_z B_r^{*} &=
    i c k A_z
    \dfrac{- i k A_z^{*}}{J_l \sin{\theta}}
    \left(
    \dfrac{l}{k_r r} J_l
    + \Lambda \cos{\theta} \dot{J}_l
    \right)
    =
    \dfrac{1}{2}
    c k^2 \sin{\theta}
    J_l
    \left(
    \dfrac{l}{k_r r} J_l
    + \Lambda \cos{\theta} \dot{J}_l
    \right), \\
    E_r B_z^{*} &=
    -\dfrac{c k}{\sqrt{2}}
    \dfrac{A_z}{d^1_{0 \Lambda}}
    \left(
    \dfrac{l}{k_r r}
    + \Lambda \cos{\theta}
    \dfrac{\dot{J}_l}{J_l}
    \right)
    \dfrac{A_z^{*} k_r}{\sqrt{2} d^1_{0 \Lambda}}
    =
    - \dfrac{1}{2}
    c k^2 \sin{\theta}
    J_l
    \left(
    \dfrac{l}{k_r r} J_l
    + \Lambda \cos{\theta} \dot{J}_l
    \right).
\end{align}

Combining these expressions, we obtain
\begin{align}
    S_{\varphi} &=
    \dfrac{c k^2 \sin{\theta}}{2 \mu_0}
    J_l
    \left(
    \dfrac{l}{k_r r} J_l
    + \Lambda \cos{\theta} \dot{J}_l
    \right) \nonumber \\
    &=
    \dfrac{c k^2}{2 \mu_0}
    \left[
    \dfrac{l}{k r}
    \left( 1 + \Lambda \cos{\theta} \right)
    J_l ^2
    -\dfrac{\Lambda}{2}
    \sin{2 \theta}
    J_l J_{l+1}
    \right].
\end{align}

\section{Evaluation of intercepted angular momentum flux}
\label{total momentum}

In the main text, we obtained
\begin{align}
    M_z =
    \dfrac{\pi}{\mu_0 k \sin^2{\theta}}
    \left[
    l \cos{\theta}\underbrace{\int\limits_0^{k_r a} J^2_l(y)\, y\, \mathrm{d} y}_{I_1}
    + \Lambda
    \underbrace{\int\limits_0^{k_r a} J_l(y)\, \dot{J}_l(y)\, y^2\, \mathrm{d} y}_{I_2}
    \right].
\end{align}

We now evaluate the integrals $I_1$ and $I_2$.
Starting from the Bessel equation,
\begin{align}
    y^2 \ddot{J}_l + y \dot{J}_l + (y^2 - l^2) J_l = 0,
\end{align}
and multiplying it by $2\dot{J}_l$, we obtain
\begin{align}
    2y^2 \dot{J}_l \ddot{J}_l
    + 2 y \dot{J}^2_l
    + 2(y^2 - l^2) \dot{J}_l J_l = 0 .
\end{align}

It follows that
\begin{align}
    \dfrac{\mathrm{d}}{\mathrm{d} y} (y^2 \dot{J}^2_l)
    &= 2y^2 \dot{J}_l \ddot{J}_l + 2 y \dot{J}^2_l, \\
    \dfrac{\mathrm{d}}{\mathrm{d} y} \big[(y^2 -l^2)J^2_l\big]
    &= 2 y J^2_l + 2(y^2 - l^2) J_l \dot{J}_l .
\end{align}

Therefore,
\begin{align}
    \dfrac{\mathrm{d}}{\mathrm{d} y}
    \left(
    y^2 \dot{J}^2_l + (y^2 -l^2)J^2_l
    \right)
    = 2y J^2_l .
\end{align}

Integrating over the interval $[0,k_r a]$, we find
\begin{align}
    I_1 =
    \int_0^{k_r a} J^2_l(y)\, y\, \mathrm{d} y
    =
    \dfrac{(k_r a)^2}{2}
    \left\{
    \dot{J}^2_l(k_r a)
    + \left[
    1 - \left(\dfrac{l}{k_r a}\right)^2
    \right]
    J^2_l(k_r a)
    \right\}.
\end{align}

Next, we evaluate $I_2$:
\begin{multline}
I_2 =
\int_0^{k_r a} J_l(y)\, \dot{J}_l(y)\, y^2\, \mathrm{d} y
=
\int_0^{k_r a}
\dfrac{\mathrm{d} J^2_l}{2}\, y^2
= \\
\left.
\dfrac{y^2 J^2_l}{2}
\right|_0^{k_r a}
-
\int_0^{k_r a} y J^2_l\, \mathrm{d} y
=
\dfrac{(k_r a)^2 J^2_l(k_r a)}{2}
-
I_1
=
\dfrac{(k_r a)^2}{2}
\left[
\left(\dfrac{l}{k_r a}\right)^2
J^2_l(k_r a)
-
\dot{J}^2_l(k_r a)
\right].
\end{multline}

Substituting $I_1$ and $I_2$ into the expression for $M_z$, we obtain
\begin{align}
    M_z &=
    \dfrac{\pi a^2 k}{2 \mu_0}
    \Bigg\{
    l \cos{\theta}
    \left[
    \dot{J}^2_l(k_r a)
    +
    \left(
    1 - \dfrac{l^2}{k^2_r a^2}
    \right)
    J^2_l(k_r a)
    \right]
    \nonumber \\
    &\hspace{1.8cm}
    +
    \Lambda
    \left[
    \left(\dfrac{l}{k_r a}\right)^2
    J^2_l(k_r a)
    -
    \dot{J}^2_l(k_r a)
    \right]
    \Bigg\} \nonumber \\
    &=
    \dfrac{a^2 \pi k}{2 \mu_0}
    \left\{
    (l \cos{\theta} -\Lambda) \dot{J}^2_l(k_r a)
    +
    \left[
    l \cos{\theta} \left(1-\dfrac{l^2}{k^2_r a^2}\right)
    +
    \Lambda
    \left(\dfrac{l}{k_r a}\right)^2
    \right]
    J^2_l(k_r a)
    \right\}.
\end{align}

Introducing the notation $x = ka$ and using the identity
\begin{align}
    \dot{J}_l (x \sin{\theta})
    =
    \dfrac{l}{x \sin{\theta}} J_l (x \sin{\theta})
    -
    J_{l+1} (x \sin{\theta}),
\end{align}
we obtain
\begin{multline}
    M_z =
    \dfrac{a \pi}{2 \mu_0} x
    \Big[
    l \cos{\theta} J^2_l (x \sin{\theta})
    -
    \dfrac{2l}{x \sin{\theta}}
    J_l (x \sin{\theta})
    J_{l+1} (x \sin{\theta})
    (l \cos{\theta} - \Lambda)
    +
    J^2_{l+1} (x \sin{\theta})
    (l \cos{\theta} - \Lambda)
    \Big] \\
    =
    \dfrac{a \pi}{2 \mu_0} x
    \Big[
    l \cos{\theta} J^2_l (x \sin{\theta})
    +
    J_{l+1} (x \sin{\theta})
    (l \cos{\theta} - \Lambda)
    \left(
    J_{l+1} (x \sin{\theta})
    -
    \dfrac{2l}{x \sin{\theta}}
    J_l (x \sin{\theta})
    \right)
    \Big].
\end{multline}

Using the relation
\begin{align}
    J_{l+1} (x \sin{\theta})
    -
    \dfrac{2l}{x \sin{\theta}}
    J_l (x \sin{\theta})
    =
    -
    J_{l-1} (x \sin{\theta}),
\end{align}
we finally arrive at
\begin{equation}
    M_z =
    \dfrac{a \pi}{2 \mu_0} x
    \left[
    l \cos{\theta} J^2_l (x \sin{\theta})
    - \left(l\cos{\theta} - \Lambda \right)
    J_{l+1} (x \sin{\theta})
    J_{l-1} (x \sin{\theta})
    \right].
\end{equation}

\section{Estimation of the vector potential amplitude $A_0$ \label{app:sec:E}}
For our system, we first calculate explicitly the axial Poynting flux of the regularized beam given by Eq. \eqref{eq:S_z}
\begin{align}
    S^{(\mathrm{BG})}_z(r) = \frac{1}{2 \mu_0} \Re(E_r B^*_\phi - E_\phi B^*_r) \; e^{-2 r^2/w^2}.
\end{align}
The substitution of field components leads to
\begin{align}
    E_r B^*_\phi &= - \dfrac{ck^2 \Lambda A_0^2}{8} 
     \left[ (1 - \Lambda \cos{\theta})^2 J^2_{l+1} - (1 + \Lambda\cos{\theta})^2 J^2_{l-1} \right], \\
     E_\phi B^*_r &= \dfrac{ck^2 \Lambda A_0^2}{8} 
     \left[ (1 - \Lambda \cos{\theta})^2 J^2_{l+1} - (1 + \Lambda\cos{\theta})^2 J^2_{l-1} \right],
\end{align}
and as a result
\begin{align}
    S^{(\mathrm{BG})}_z(r) = \frac{c k^2 \Lambda A^2_0}{8 \mu_0} \left[(1 + \Lambda\cos{\theta})^2 J^2_{l-1} - (1 - \Lambda \cos{\theta})^2 J^2_{l+1} \right] \; e^{-2 r^2/w^2}.
\end{align}
The total incident power for a finite-energy Bessel-Gauss regulatization according to Eq. \eqref{eq:Power} is equal to
\begin{align}
    P &= 2\pi \int\limits_{0}^{\infty} \langle S^{(\mathrm{BG})}_{z}(r)\rangle rdr \nonumber \\
    &= \frac{2 \pi c k^2 \Lambda A^2_0}{8 \mu_0 k^2_r} \left[ (1 + \Lambda\cos{\theta})^2 \int \limits^\infty_0 J^2_{l-1} (y) e^{-2 y^2/(k^2_r w^2)} y dy - (1 - \Lambda \cos{\theta})^2 \int \limits^\infty_0 J^2_{l+1} (y) e^{-2 y^2/(k^2_r w^2)} y dy \right].
\end{align}
The analytical answer can be found with the help of expression 6.615 in the book \cite{GR}
\begin{align}
    \int \limits^\infty_0 J^2_l (\sqrt{z}) e^{- \alpha z} dz = \frac{1}{\alpha} I_l\left(\frac{1}{2 \alpha}\right) e^{-1/(2\alpha)},
\end{align}
where $I_l$ is the modified Bessel function of the first kind.

We finally obtain the total incident power for a finite-energy Bessel-Gauss regulatization in the form
\begin{align}
    P = \frac{\pi c k^2 w^2 \Lambda A^2_0}{16 \mu_0} e^{- k^2 w^2 \sin^2{\theta} / 4} \left[ (1 + \Lambda\cos{\theta})^2 I_{l-1} \left(\frac{k^2 w^2 \sin^2{\theta}}{4} \right) - (1 - \Lambda \cos{\theta})^2 I_{l+1} \left(\frac{k^2 w^2 \sin^2{\theta}}{4} \right) \right].
\end{align}
The amplitude of the vector potential is
\begin{align}
    A_0 = \sqrt{\frac{16 \mu_0 P}{\pi c k^2 w^2 \Lambda}} e^{k^2 w^2 \sin^2{\theta} / 8} \left[ (1 + \Lambda\cos{\theta})^2 I_{l-1} \left(\frac{k^2 w^2 \sin^2{\theta}}{4} \right) - (1 - \Lambda \cos{\theta})^2 I_{l+1} \left(\frac{k^2 w^2 \sin^2{\theta}}{4} \right) \right]^{-1/2}.
\end{align}
For the incident paraxial beam with an opening angle $\theta = 0.01$ rad, wavelength $\lambda = 532$ nm, polarization $\Lambda = 1$, $l=1$, power $1$ kW, and the transverse envelope scale $w = 5$ cm the vector potential amplitude takes the value of
\begin{align}
    A_0 \approx 3.4 \cdot 10^{-10} \text{ T} \cdot \text{m}.
\end{align}
\bibliographystyle{apsrev4-2} 
\bibliography{references}  
\end{document}